\documentstyle[preprint,aps]{revtex}
\begin{document}
\title{Microscopic origin of the conducting channels 
in metallic atomic-size contacts}

\author{J.C. Cuevas, A. Levy Yeyati and A. Mart\'{\i}n-Rodero}

\address{Departamento de F\'\i sica Teo\'rica de la Materia Condensada C-V, 
Universidad Aut\'onoma de Madrid, E-28049 Madrid, Spain}

\date{\today}

\maketitle

\begin{abstract}
We present a theoretical approach which allows to determine
the number and orbital character of the conducting channels in metallic
atomic contacts. We show how the conducting channels
arise from the atomic orbitals having a
significant contribution to the bands around the Fermi level. 
Our theory predicts that the number
of conducting channels with non negligible transmission is 3 for Al and
5 for Nb one-atom contacts, in agreement with recent experiments. 
These results are shown to be robust with respect to 
disorder. The experimental values of the channels transmissions lie 
within the calculated distributions. 
\end{abstract}



\narrowtext

Metallic contacts of atomic dimensions, that can be produced  
by means of scanning tunneling microscope and break-junctions
techniques, have been the object of great attention in the last few
years \cite{review}. According to the scattering approach,
the electronic transport
in these mesoscopic structures can be described 
in terms of independent ``conducting channels'' characterized by
certain transmission coefficients which vary between zero and one 
\cite{Landauer}. 
The complete understanding of their transport properties requires 
the knowledge of the transmission coefficients along {\it each} conducting 
channel. 
This information, which is not accessible by usual conductance
measurements, has for the first time been obtained in recent experiments
on superconducting Al contacts \cite{Scheer}.
The possibility of extracting the individual transmissions
from measuring the superconducting $I-V$ characteristic is based on the 
extreme sensitivity of the subgap structure to small changes in these
parameters \cite{us,Averin}. 

A remarkable conclusion of this experimental study is that,
although the total conductance of an Al contact on the first plateau
(pressumably a one-atom contact \cite{Krans}) can be close
to one quanta of conductance, this situation does not 
correspond to a nearly open single channel, but
rather to a situation with three partially open channels. 
The question naturally arises on the microscopic origin of this
phenomena for the specific case of Al. More generally, one would like to
know what is the number of conducting channels for a given metal in a
given contact geometry.
The aim of this letter is to provide theoretical insight
into these questions. 

Having a system of atomic dimensions
it seems natural to choose an atomic orbital basis for the
description of its electronic structure. 
This choice has proven useful in the context of STM 
theory \cite{STM2}. Furthermore, the use of a local basis in combination
with Green function techniques provides an efficient way for obtaining
the transport properties \cite{us,STM2,Sutton} in terms of microscopic
parameters. 
In an atomic orbital basis the electronic Hamiltonian adopts the 
usual tight-binding form

\begin{equation}
\hat{H} =  \sum_{i \alpha, \sigma} \epsilon_{i \alpha}
c^{\dagger}_{i \alpha \sigma} c_{i \alpha \sigma}
+ \sum_{i \alpha \neq j \beta, \sigma} t_{i\alpha,j\beta}
c^{\dagger}_{i \alpha,\sigma}
c_{j \beta,\sigma} ,
\end{equation}

\noindent
where $i,j$ run over the atomic sites and $\alpha,\beta$ denote the
different atomic orbitals (the number of orbitals at each site will be 
denoted by $N_{orb}$). The hopping elements  $t_{i\alpha, j\beta}$ are
assumed to connect first-neighbor sites only.
There exist in the literature several parametrization procedures for
determining the tight-binding Hamitonian \cite{Harri,Papa} which are
known to accurately reproduce the band structure of bulk materials.
As a starting point we follow the parametrization proposed in Ref. 
\cite{Papa} considering as a minimal basis for each metal those atomic
orbitals having a significant contribution to the electronic DOS around
the Fermi energy, $E_F$. Thus, for the case of superconducting metals
of groups III and IV, like Al, Pb, etc, only $s$ and $p$ orbitals need
to be considered; while for transition metals like Nb $d$ orbitals have
to be included.

In an atomic contact the local environment in the neck region is very
different to that of the bulk material and therefore the use of bulk
parameters in the Hamiltonian requires some justification. In the first
place, the inhomogeneity of the contact geometry can produce large deviations
from the approximate local charge neutrality that typical metallic
elements must exhibit. Within the tight-binding
approximation this effect can be corrected imposing local charge neutrality 
through a self-consistent variation of the diagonal parameters $\epsilon_{i
\alpha}$ \cite{us2}. As discussed below, this self-consistency in the
neck region turns out to be crucial for the correct determination of
the conducting channels. Regarding the hopping elements, $t_{
i\alpha,j\beta}$, although we shall initially consider them as being equal to
the bulk values in order to represent a neck geometry with bulk
interatomic distances; we shall show that the results are robust with
respect to fluctuations in the hopping elements induced by disorder in
the atomic positions.

An idealized geometry for a one-atom contact is depicted in Fig.1. It
consists of a close-packed fcc structure grown along
the (111) direction (hereafter denoted as z direction), starting from
a central atom \cite{simulations}. 
This structure is connected to two semi-infinite crystals
describing the metallic leads, $N$ being the number of atomic layers
within the neck region. By taking different values of $N$ 
we can describe both the cases of long and short necks.

In order to obtain the d.c. current for a constant bias voltage 
applied between the leads, it is most convenient to use non-equilibrium
Green function techniques \cite{Caroli}. Within this formalism the current can
be written with an expression formally equivalent to that of scattering
theory \cite{Nazarov} by treating the coupling between the central
region and the leads as a perturbation \cite{aly}. The current
between the left lead and the central region is then given by

\begin{equation}
I = \frac{2e}{h} \int^{\infty}_{-\infty} T(E,V)
\left [ f_L(E) - f_R(E) \right] dE ,
\end{equation}

\noindent
where $f_{L,R}$ are the Fermi-distribution functions for the (left, right)
leads and $T(E,V)$ is an energy  and voltage dependent
transmission probability, which can be written in terms of 
the matrix elements of the (retarded, advanced) Green function
matrix $\hat{G}^{r,a}(E) = [ E \pm i0^{+} - \hat{H} ]^{-1}$ as

\begin{eqnarray} T(E,V) & = & 4 \mbox{Tr} \left[\mbox{Im}
\hat{\Sigma}_L(E-e\frac{V}{2}) \right. \times \nonumber \\
& & \left. \hat{G}^r_{1N}(E) 
\mbox{Im} \hat{\Sigma}_R(E+e\frac{V}{2}) \hat{G}^a_{N1}(E) \right]. \nonumber 
\end{eqnarray}

In this expression 
$\hat{G}^{r}_{1N}$ and $\hat{G}^{a}_{N1}$ are matrices whose elements
are the Green functions connecting layers 1 and $N$, 
$\hat{\Sigma}_{L,R}$ being self-energy matrices describing
the coupling of the central region to the leads. These
matrices have a dimension equal to the number of bonds connecting the 
central region to the leads ($M_{L,R}$) and have a simple expression in
terms of the Green functions of the uncoupled leads, $g_{i\alpha,j\beta}$:        

\[ \left(\hat{\Sigma}_{L,R}(E) \right)_{i\alpha,j\beta} = 
\sum_{k,l \in L,R \; ; \; \gamma,\delta} t_{i\alpha,k\gamma} 
g^a_{k\gamma,l\delta}(E) t_{l\delta,j\beta}. \] 

The $g_{i\alpha,j\beta}$'s can be evaluated numerically by standard 
decimation techniques \cite{decimation}. 

The voltage range which is probed in the experiments with
superconducting contacts is of the order of the gap parameter 
$\Delta$ (typically a few tenths of $meV$) \cite{Scheer}. 
For this bias range normal metallic
systems will behave ohmically. Even when the atomic-size contacts exhibit
resonances around $E_F$ \cite{us2}, their width will in general
be much larger than
$\Delta$ and linearization of Eq. (2) is appropriate.
In this linear regime the contact normal conductance can be expressed as
$G = \left(2e^2/h\right) T(E_F,0)$. By using the cyclic property of
the trace,  $G$ can be written in the Landauer form $G = \left(2e^2/h
\right) \mbox{Tr}[\hat{t}(E_F) \hat{t}^{\dagger}(E_F)]$, where 

\begin{equation}
\hat{t}(E) = 2 \left[ \mbox{Im} \hat{\Sigma}_L(E) \right]^{1/2} 
\hat{G}^r_{1N}(E) \left[ \mbox{Im} \hat{\Sigma}_R(E) \right]^{1/2}.
\end{equation}

The existence of $(\mbox{Im} \hat{\Sigma} )^{1/2}$ 
as a real matrix is warranted by $\mbox{Im} \hat{\Sigma}$
being positive definite. Moreover, $\hat{t} \hat{t}^{\dagger}$ is an
hermitian matrix having $M_L$ real eigenvalues, $T_i$, which are bounded
by zero and one \cite{comment3}. Associated with these eigenvalues
there will be $M_L$ eigenvectors, which in our model are linear
combinations of the atomic orbitals in the layer which is in
contact with the left lead. These eigenvectors define the way in which 
the atomic orbitals contribute to each conducting channel. 

Although the dimension of $\hat{t} \hat{t}^{\dagger}$ can be arbitrarily 
large depending on the size of the central region, the actual number of
conducting channels (those with a non-vanishing transmission eigenvalue
$T_i$) are limited by the number of orbitals in the narrowest section of
the neck ($N_{orb}$ when having a single atom contact). 
This fact can be shown by the following simple argument. As the division
between ``central region'' and leads is somewhat arbitrary, one could
always redefine the leads for the geometry of Fig. 1 in such a way that
the new central region would only consist of the central atom. Then, the
new lead self-energy matrices $\hat{\Sigma}^{\prime}_{L,R}$
would have a dimension of just $N_{orb}$ and the new transmission matrix
would only admit $N_{orb}$ eigenmodes. Current conservation along each
conducting channel ensures that the nonvanishing eigenvalues $T_i$ and
$T^{\prime}_i$ must be the same. 

The above simple argument already allows an estimate of the maximum
number of relevant conducting channels in a one-atom contact.
Thus, for an $sp$-like metal like Al, this number
should be typically four, while for a transition metal like Nb (having a
negligible weight of $p$ orbitals at $E_F$) this number would be of 
order six. As discussed below, this rough estimate should be taken as
an upper bound. The actual number of conducting channels can be smaller
as some of the channels can carry no current due to symmetry
considerations. 

Let us first consider the case of an Al one-atom contact, which is the
one analyzed experimentally in Ref. \cite{Scheer}. Al contacts have also
been addressed theoretically in \cite{Lang}. The atomic
configuration for Al, $3s^23p^1$, gives rise to a conducting $sp$ band
with three electrons per atom. The bulk DOS at $E_F$ has 
important contributions from both $3s$ and $3p$ orbitals, the $3s$
level being located $\sim 7 eV$ below the $3p$ level \cite{Papa}. 
While the above simple argument would predict a maximum number of
four channels for a one-atom contact, the self-consistent calculation
for the ideal geometry yields only three channels with
nonvanishing transmission. In Fig. 2 the transmission eigenvalues of the
ideal (111) contact are shown as a function of energy for the two extreme
cases of a short ($N=1$) or a long ($N \rightarrow \infty$) neck
geometry. There are certain features that are common
to both cases: i) there are three channels having a significant
transmission around $E_F$, the fourth one being closed for every energy;
ii) the total transmission is close to one around $E_F$ increasing
to almost three at higher energies; iii) there is a non-degenerate mode
which is widely open for almost every energy; iv) the second
transmission eigenvalue is two-fold degenerate and has a small value
around $E_F$. 

The channels can be classified according to the orbital character of the
eigenvectors on the central atom. In this way, the non-degenerate channel, 
widely open at $E_F$, is a combination with an amplitude $\sim 0.63$ on the
$s$ and $\sim 0.77$ on the $p$ orbitals
along the $z$ direction. The one with zero transmission corresponds
to the orthogonal combination of these two orbitals. On the other hand, 
the two degenerate modes correspond
to combinations of $p$ orbitals on a plane perpendicular to the $z$
direction. While the symmetry properties of the neck geometry are
responsible for the decoupling between the $s-p_z$ and $p_x-p_y$
orbitals, the approximate fulfillment of the condition $\mbox{Im}
\Sigma^{\prime}_{ss} \mbox{Im}\Sigma^{\prime}_{p_z p_z} = 
\left(\mbox{Im}\Sigma^{\prime}_{s p_z}\right)^2$ accounts for the
presence of a non-conducting channel (details will be given elsewhere).

The orbital character and 
energy dependence of the transmission eigenvalues is similar for the
case of Pb, which is also an $sp$-like metal. However in this case the
Fermi level lies in the region where the three channels are almost open,
giving rise to a total transmission larger than 2.   

So far the possible effect of disorder in the atomic positions has
been disregarded. We have studied this effect by introducing random
fluctuations in the atomic positions of the
idealized structure, assuming the distance dependence on the hopping
parameters as suggested by Harrison \cite{Harri}. Although there are
certain features like the two-fold degeneracy which, as expected,
disappear with the inclusion of disorder, the gross features found for
the ideal geometry are nevertheless robust. This fact is illustrated in the
histograms shown in Fig. 3, which demonstrate that the
decomposition of the total transmission consists of a widely open
channel with $T_i$ between 0.6 and 0.9 and two low 
transmissive channels with $T_i$ between 0.1 and 0.3. These predictions
are consistent with the experimental results for the first
conductance plateau \cite{Scheer}. The fourth channel always
has an extremely small transmission ($T_i < 10^{-4}$). 

We have also analyzed the conducting channels of a Nb one-atom contact,
as an example of transition metal contacts. 
As commented above, in this
case the maximum number of conducting channels is expected to be six due
to the fact that the DOS around $E_F$ mainly arises from the contributions
of $5s$ and $4d$ orbitals \cite{Papa}. 
The idealized one-atom contact
geometry yields in this case a total transmission between 2 and 3
depending on the number of layers in the central region. 
The channel decomposition shows that this total transmission is mainly built  
up from the contribution of five conducting channels. 
As can be observed in Fig. 4,
the $d$ bands cause a strong energy dependence of the
transmission eigenvalues, with
typical energy scales of the order of $0.5 eV$. The $s$
and $d_{z^2}$ orbitals hybridize strongly and give rise to the
conducting channel with the highest transmission around $E_F$ (mode 1
in Fig. 4). 
The almost closed channel (mode 6 in Fig. 4) corresponds to the 
orthogonal combination of these two orbitals. There also
appears a two-fold degenerate channel with transmission $\sim 0.7$
and another two-fold degenerate channel with transmission $\sim 0.3$. 
Both the values of the total transmission and the number of relevant
conducting channels are in good agreement with preliminary experimental
results \cite{Ludoph}.

Our theory and the experimental results show that atomic contacts
of metals which have an important
contribution from $p$ and $d$ orbitals, do not necessarily, even in the
one-atom case, exhibit an integer number of
perfectly transmitting modes. This situation is at variance with that 
of simple metals like Na, Au, Ag, etc, which can be described by a single 
$s$ like band around
$E_F$ and exhibit well defined quantized conductance steps, at least
at the lower plateaus \cite{steps}. Within our theory, in the one-atom
contact geometry, simple metals would have a single conducting channel.
The transmission of that channel is strongly pinned at one due to
the charge neutrality condition \cite{us2}.

In conclusion, we have presented a theoretical analysis of the
conducting channels in metallic atomic-size contacts. We have
shown that the number and character of these channels
are determined by the orbital electronic structure and the local atomic
environment around the neck region.  
For the case of $sp$-like metals like Al and Pb we predict the presence
of 3 conducting channels in a one-atom contact, in good agreement with
the available experimental data \cite{Scheer}.
For one-atom contacts of a transition metal like Nb we expect
the presence of 5 conducting channels due to the contribution of $d$
orbitals. This result has been confirmed by recent experiments
\cite{Ludoph}.

\acknowledgements
The authors would like to thank E. Sheer, C. Urbina, J.M. van Ruitenbeek
and
N. Agra\"{\i}t for very fruitful discussions and for showing us some of their
experimental results before publication. Useful remarks from F. Flores
and F.J. Garc\'{\i}a-Vidal are also acknowledged. 
This work was partially supported by the Spanish CICYT 
(contract No. PB93-0260).

\begin{figure}[h]
\caption[]
{\label{fig1} Idealized geometry for a one-atom contact. The layers are
numbered from 1 to $N$ starting from the left lead.}
\end{figure}

\begin{figure}[h]
\caption[]
{\label{fig2} Transmission eigenvalues as a function of energy for
Al one-atom contacts in the two extreme cases of short (a) and long
(b) necks. The solid curve corresponds to the non-degenerate $sp_z$ mode
and the dotted curve corresponds to the two-fold degenerate $p_x-p_y$ mode.} 
\end{figure}

\begin{figure}[thbp]
\caption[]
{\label{fig3} Typical distributions of the transmission eigenvalues for 
an Al one-atom contact when disorder in the atomic positions is
included (the maximum fluctuation in the hopping parameters is of the 
order of 100$\%$ with respect to the bulk values, which corresponds to
variations of the order of 20$\%$ in the interatomic distances).
The two modes corresponding to the two-fold degenerate eigenvalue in the
idealized case exhibit a similar distribution.}
\end{figure}

\begin{figure}[h]
\caption[]
{\label{fig4}
Transmission eigenvalues as a function of energy for a Nb one-atom
contact for the long neck case. 
Note the two-fold degeneracy of modes 2-3 and of 4-5.}
\end{figure}


\begin{references}
\bibitem{comment3} These bounds can be derived using the general
properties of the Green functions.
\end{references}
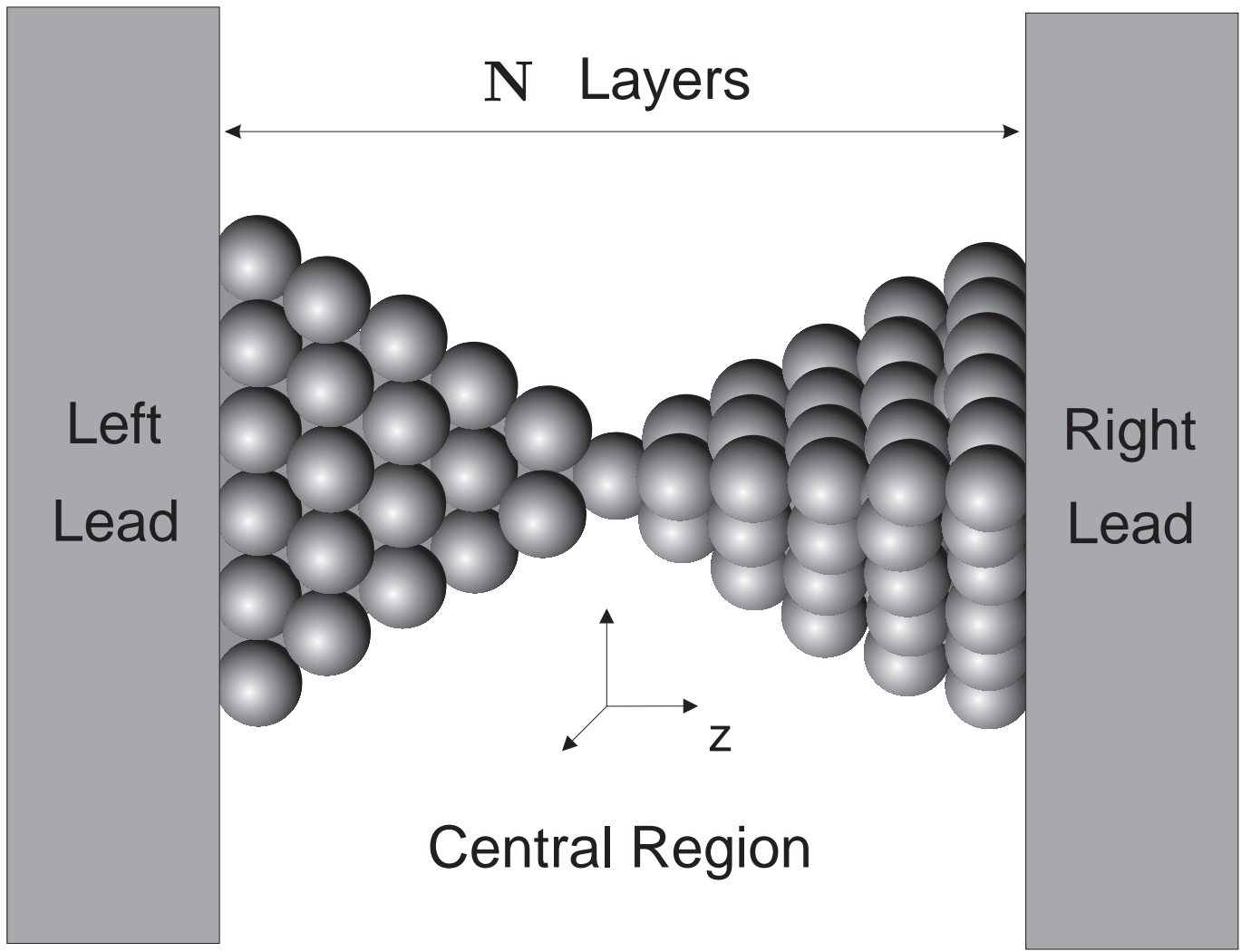
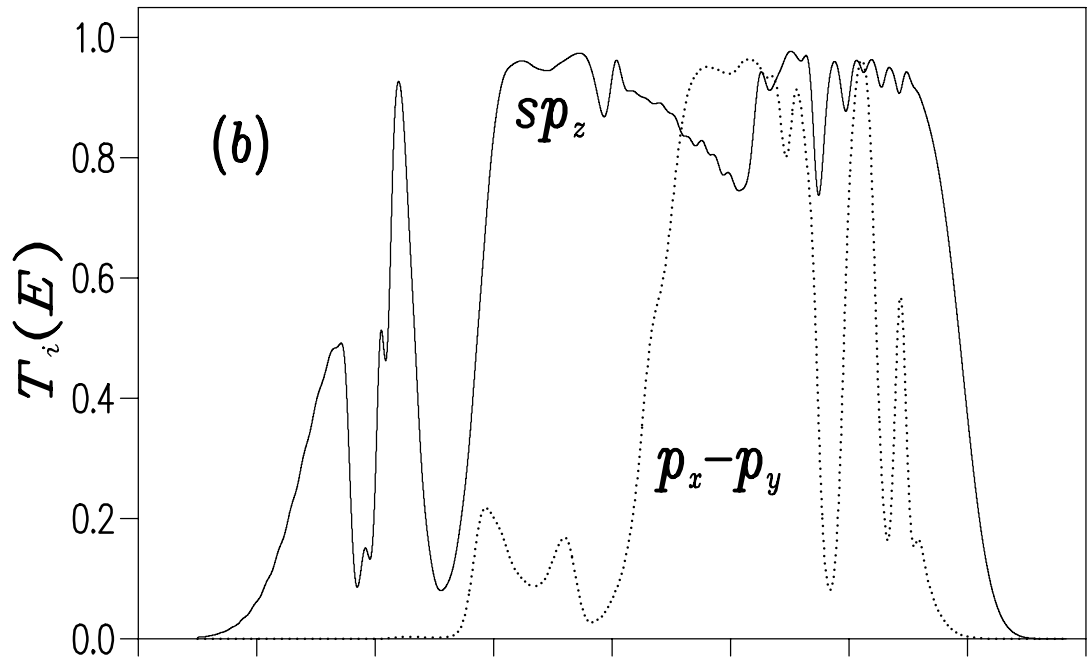
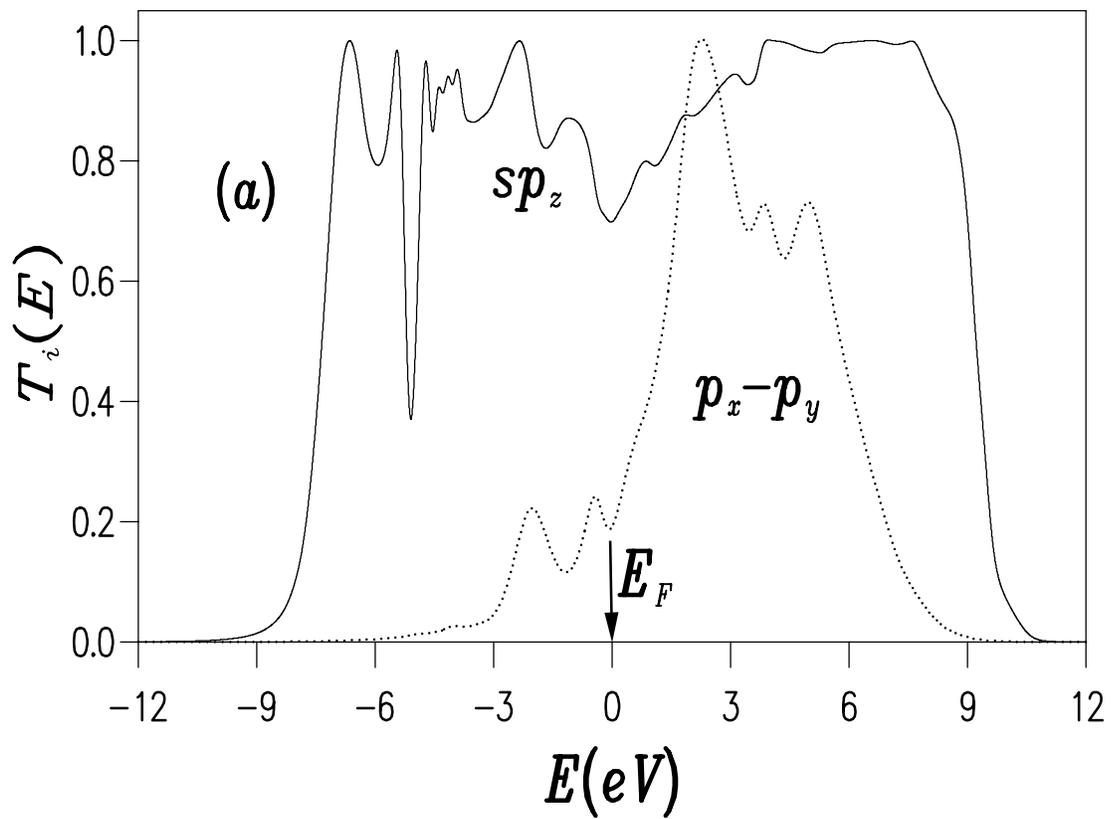
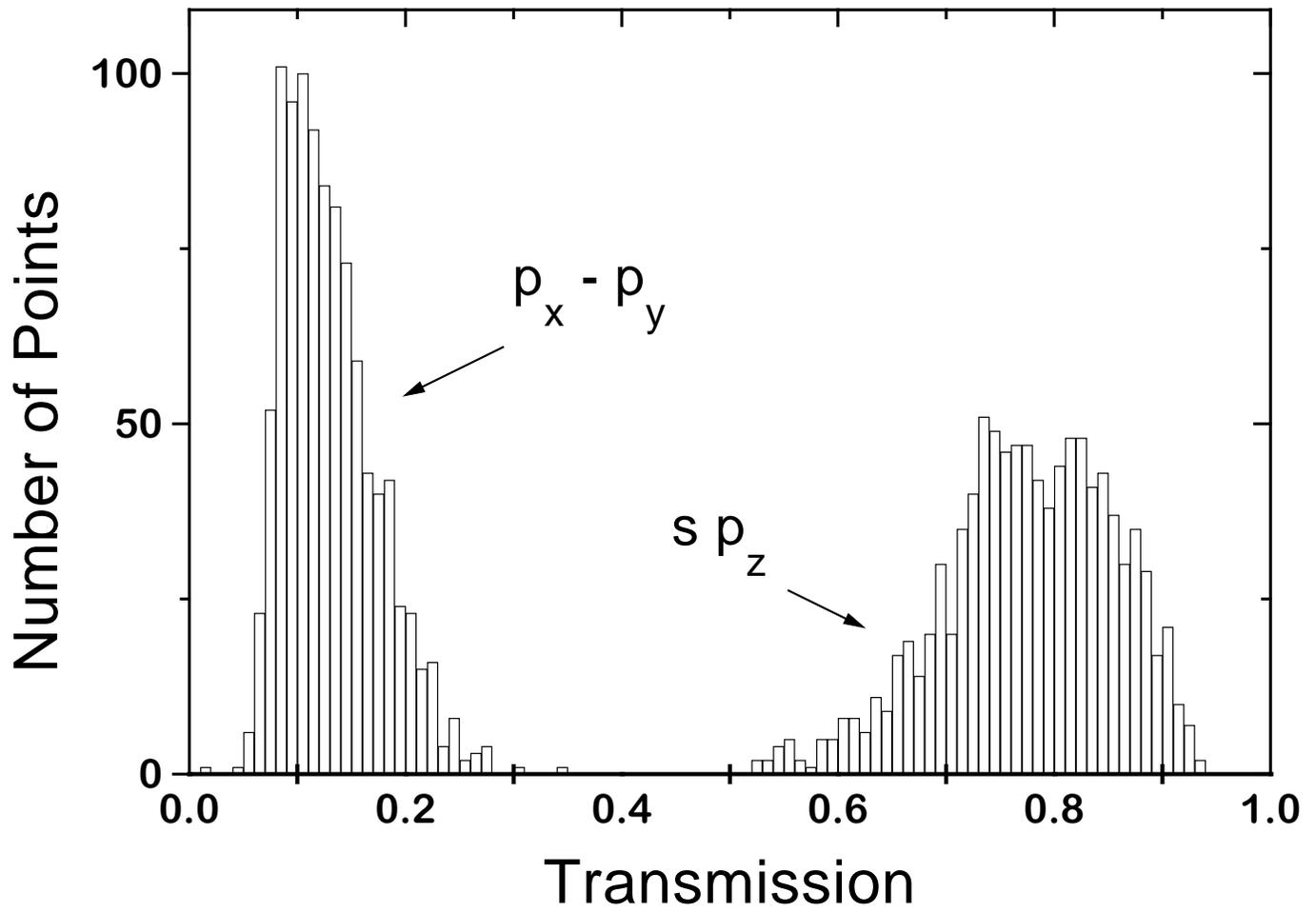
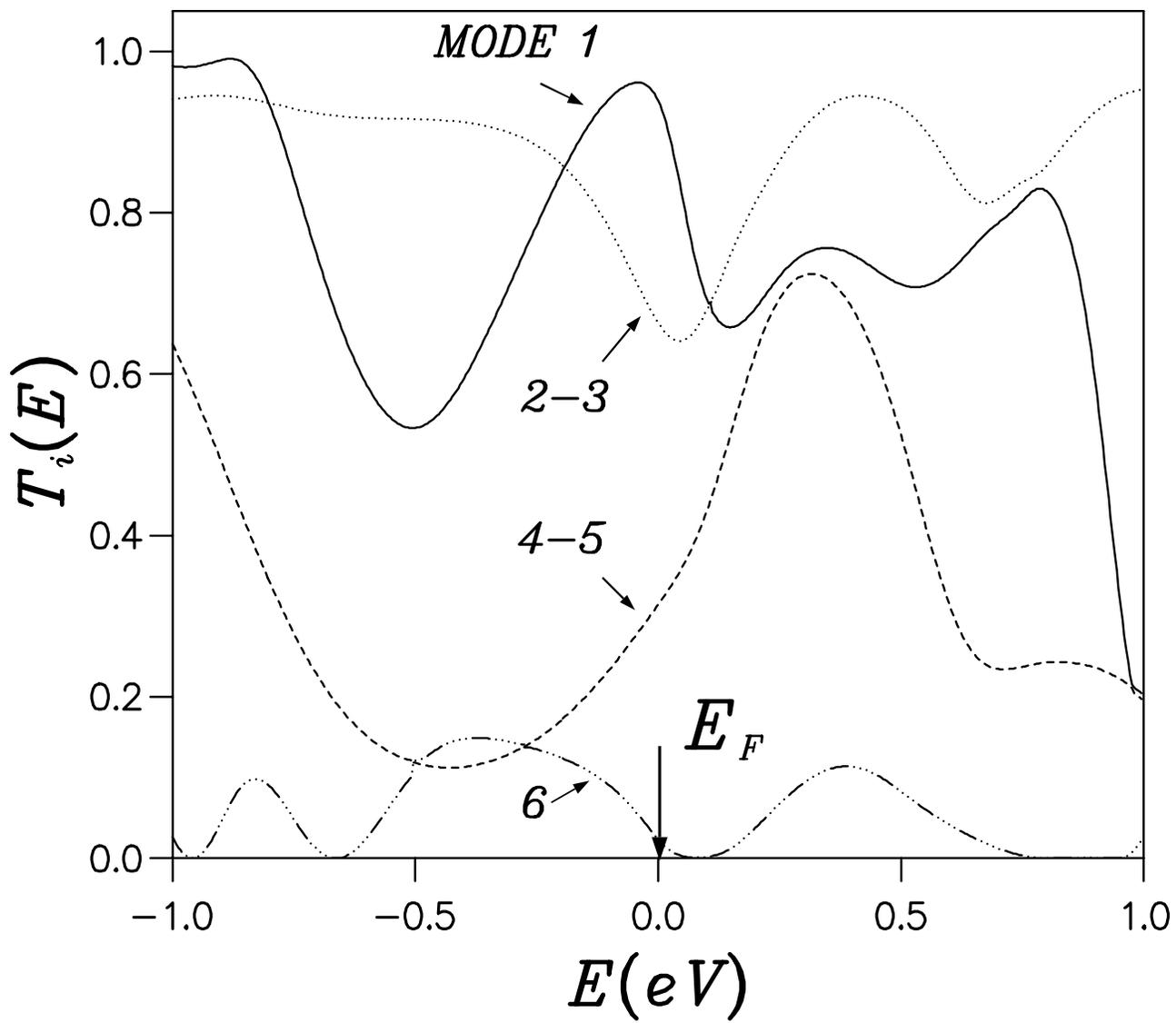
\end{document}